# Using Arabic Tweets to Understand Drug Selling Behaviors

Wesam Alruwaili[a], Bradley Protano[b], Tejasvi Sirigiriraju[c], Hamed Alhoori[d]

*[a]Department of Computer Science, Jouf University, Sakaka 42421, Saudi Arabia*
*[b]Department of Computer Science, Northern Illinois University, Dekalb 60115, United State*
*[c]Department of Computer Science, Northern Illinois University, Dekalb 60115, United State*
*[d]Department of Computer Science, Northern Illinois University, Dekalb 60115, United State*

**Abstract**

Twitter is a popular platform for e-commerce in the Arab region—including the sale of illegal goods and services. Social media platforms present multiple opportunities to mine information about behaviors pertaining to both illicit and pharmaceutical drugs and likewise to legal prescription drugs sold without a prescription, i.e., illegally. Recognized as a public health risk, the sale and use of illegal drugs, counterfeit versions of legal drugs, and legal drugs sold without a prescription constitute a widespread problem that is reflected in and facilitated by social media. Twitter provides a crucial resource for monitoring legal and illegal drug sales in order to support the larger goal of finding ways to protect patient safety. We collected our dataset using Arabic keywords. We then categorized the data using four machine learning classifiers. Based on a comparison of the respective results, we assessed the accuracy of each classifier in predicting two important considerations in analysing the extent to which drugs are available on social media: references to drugs for sale and the legality/illegality of the drugs thus advertised. For predicting tweets selling drugs, Support Vector Machine, yielded the highest accuracy rate (96%), whereas for predicting the legality of the advertised drugs, the Naïve Bayes, classifier yielded the highest accuracy rate (85%).

*Keywords:* Public Health; Twitter; Social Media; Drug; Drug Seller; Arabic Language; Pharmacological.

## 1. Introduction

Twitter is accessible to most of the world, including the Arab region. Users post tweets to share news of various kinds with others and for such purposes such as self-promotion and sales—including the sale of drugs without oversight. Given the health risks associated with taking drugs under these conditions, we collected data to monitor the drug trade on Twitter.

It is certainly the case that some of the drugs available through Twitter are authentic medications, legally sold. However, illegal drugs, legal but controlled drugs sold without a prescription and/or in illegal dosages, and counterfeits of legal drugs are all readily available on this platform. In addition, these drugs, even those that can be sold legally elsewhere, carry further risks when obtained in this context as many sellers on Twitter may not be aware of the importance of storing and transporting drugs properly or even of the importance of expiration dates. Or, they may not care about such matters. Either way, it is certain that compromised, expired, illicit, fake, and stolen drugs—all of which endanger human life—are being sold on the Twitter platform. Several researchers have worked on mining English-language tweets with the goal of monitoring the buying and selling of drugs on Twitter. However, there is very little research on Arabic-language tweets of this nature due to the complexities of the language.

In this paper, we classify Arabic tweets using machine learning algorithms. We compare the results of four classifiers for our Twitter data. We address the following research questions: (1) How can we determine whether the content of a tweet is related to selling a drug, providing a health tip, or telling a joke? (2) How can we determine whether a tweet is focused on selling legal or illegal drugs?

## 2. Related Work

Several research studies have shown that microblogging platforms such as Twitter can be a useful resource for analyzing data in order to further the overall goal of protecting public health. For example, using Twitter, Buntain and Golbeck [1] showed regional drug popularity trends across the United States. Using geolocation information to determine the most and least popular drugs (e.g., cocaine, heroin, methampetamine, and oxycontin) according to region, the researchers were able to obtain results more quickly than by using conventional methods. Also using Twitter, Sarker et al. [2] departed from the traditional research focus on drug makers and providers to consider the demographics of drug abusers and their patterns. They used supervised classification to identify abuse tweets and non-abuse tweets for four drugs (Adderall, oxycodone, quetiapine, and metformin).

Some of the richest resources relevant to public health issues are found in English-language texts, whereas resources in other languages, including Arabic, are very limited. Al-Ibrahim, Al- Khalifa, and Al-Salman [3] built an Arabic drug corpus composed of 202 drugs, each of which is described according to its generic name, brand name, chemical formula, and class. Building on previous work, Abozinadah, Mbaziira, and Jones [4] used several machine learning algorithms to detect Twitter accounts posting information in Arabic in order to market illegal products and/or provide information about how to obtain such products. According to their results, the Naïve Bayes classifier performed better than either the Support Vector Machine or the Decision Tree (J48) classifier in detecting tweets offering the target information.

A number of studies have shown that illegal drug sales are a regular occurrence on Twitter. Seaman and Giraud-Carrier [5] used a set of names for 73 well-known drugs, both legal and illegal, to monitor the words most commonly associated with the sale of drugs. They found that tweets on the subject of drugs are prevalent on Twitter and that most of these deal with illegal substances. Further, the researchers found that tweets about buying and selling illegal drugs appeared more often than those dealing with their legal counterparts. Other studies on microblog retrieval focus on pharmaceutical spam, whereby spammers advertise pharmaceuticals, such as Viagra, Levitra, and Xanax, for sale. Using mining techniques to classify tweets, Shekar, Liszka, and Chan [6] identified primary and secondary words (e.g., Viagra, Xanax, pharma, prescription, pill, medication, generic, refill) to detect true pharmaceutical spam found in tweets.

Several studies show the adverse effects of drug use, thereby confirming that monitoring the sale of drugs is important. Bian, Topaloglu, and Yu [7] used machine learning techniques to identify adverse drug-related events mentioned in tweets. They developed a classifier to identify the side effects for each drug of interest. Several recent studies such as that by Yates, Goharian, and Frieder [8] focus on exploring social media in order to detect previously unknown adverse drug reactions that are not listed on a drug's label. Sarker and Gonzalez [9] applied natural language processing and machine learning methods to automatically classify social media texts with the goal of detecting mentions of adverse drug reactions and non-adverse drug reactions. They focused on extracting tweets that referenced drugs prescribed for chronic diseases and conditions. Similarly, Plachouras, Leidner, and Garrow [10] used a binary classifier to identify tweets referencing adverse events. The researchers monitored the most frequent topics discussed in tweets in which pharmaceutical drugs are mentioned. They found adverse events to be the most discussed topic followed by the efficacy of the drugs. By using patient comments posted to social media, Rastegar-Mojarad, Liu, and Nambisan [11] identified drug-repositioning candidates, i.e., drugs with established uses that may be useful in effectively treating new diseases. The researchers developed processes to capture both the benefits and adverse effects referenced in tweets.

In some studies, researchers have analyzed activity on Twitter in order to track diseases and other ailments. Sidana et al. [12], for example, identified health-related tweets and non-health-related tweets, modeled the development of ailments, and predicted diseases. Zhang et al. [13] developed a model that uses several data sources (i.e., data on hospital visits, the prevalence of asthma among adults in the US, and Twitter) to monitor chronic diseases, including asthma, and to predict hospital visits. Byrd, Mansurov, and Baysal [14] demonstrated how to use tweets to detect and surveil influenza in a given area at a given time. They obtained a high rate of accuracy in predicting the spread of influenza. Similarly, Imran and Castillo [15] used Artificial Intelligence for Disaster Response (AIDR) to monitor flu-related tweets. Lee, Agrawal, and Choudhary [16] also used Twitter data to build a system for disease surveillance, specifically to track the spread of flu and cancer in the United States. Moreover, they monitored the types, symptoms, and treatments of these diseases.

Choudhury [17] used social media data to measure and survey mental health. Choudhury, Counts, and Horvitz [18] used Twitter data to capture depression patterns using emotion and time features by building a classifier to predict whether a tweet could be depression-indicative. They measured the level of depression in populations across geography and gender. Smrz and Otrusina [19] analyzed social media content in order to determine its usefulness in predicting epidemiological events and identified the pros and cons of using Twitter for this kind of analysis. Denecke, Dolog, and Smrz [20] created a system, referred to as the M-Eco project, that uses tweets to both detect emerging health threats early and alert users to those threats. Dumbrell and Steele [21] analyzed tweets and retweets offering public health advice to support Australian health organizations in monitoring the dissemination of health information across Twitter.

Taken together, these studies establish the effects of drug use, filter and catalogue tweets, and demonstrate ways to track drug use and predict and monitor the health consequences of illegal drugs. This body of research, therefore, constitutes a foundation for the present paper in which we build on these findings and show how they can be considered together in our study.

## 3. Data and Methods

### 3.1. Data collection and extraction

We used Twitter's streaming API to collect 200,000 Arabic tweets for the time period of October 1 to December 31, 2016. Tweets were identified and collected by tracking specific Arabic keywords— "علاج", "حبوب", " بيع حبوب", "أدوية", "كبسولات", "حبوب للبيع" translated as "medication," "drugs," "capsules," "sell drugs," "drug sale," and "treatment," respectively. Twitter users refer to drugs in their tweets by translating the English drug name into Arabic or writing the English name of the drug in Arabic. For instance, for a drug called "Nature's Bounty Capsules," some users tweet "كبسولات ناتشر باونتي" and others tweet "كبسولات عطاء الطبيعة".

First, we filtered the dataset by removing symbols (i.e., RT, @, and emojis) and extracting tweet-text duplicates. Second, we excluded tweets from Twitter-verified accounts, as it is unlikely such accounts would be used to sell fake, counterfeit, or unregistered drugs. Following this preprocessing, the dataset contained a total of 2,000 tweets. Finally, we extracted two sets of features: (1) account features, i.e., user ID, username, screen name, location, and verified/unverified account, and (2) tweet features, i.e., tweet id, tweet time, and tweet text.

### 3.2. Data translation

The texts used as data were translated into English (Table 1) by three translators who were given no information about our study or analysis in order to avoid bias in the translation. We engaged translators because automatic translation would not have been effective given that most of the tweets were written in hard-to-translate informal Arabic language and various Arabic dialects. For example, the Arabic keyword, "حبوب" can be translated into English as "pills," "acne," "grains," or "lovable." Therefore, every tweet was manually vetted to determine its context and thus the meaning of given words.

Table 1. Examples of original and translated tweets

| Original Tweet | English Translation |
|---|---|
| متوفر لدينا الان سايتوتيك الاصلي لاجهاض الحمل غير المرغوب به في اشهره الاولى بامان. | The original Cytotec is now available for a safe abortion in the first months. |
| حبوب الخميرة فقط ٧٩ ريال xxxxxxxx للطلب واتس اب #الخميرة #حبوب_الخميرة #الخميرة_الأمريكي تسليم باليد في الرياض وجدة. | Yeast pills. The price is 79 Saudi Arabian Riyals. WhatsApp number: xxxxxxxx #Yeast_pills #Yeast #American_Yeast. Hand-delivery to Riyadh and Jeddah. |
| دكتور ممكن اسال عن حبوب القمح ممكن رقم جوال الاردن للتواصل | Doctor, can I ask you about wheat grain. I need your Jordanian phone number. |

*3.3. Classification Method*

*3.3.1 Detecting drug selling.* We found that any given tweet referencing a drug could be offering a health tip, functioning as spam, or telling a joke. Therefore, we labeled each tweet manually: "yes" if selling a drug or "no" otherwise (Table 2). Then, we split the dataset into a 70% training set and a 30% test set. We applied four classifiers: Support Vector Machine (SVM), Decision Tree (DT), Naïve Bayes (NB), and Random Forest (RF).

Table 2. Examples of tweets with yes/no class labels

| Tweet | Label |
|---|---|
| Collagen tablets with direct shipment to Amman. Use discount code NHR615 #Iherb. | Yes |
| The side effects of Cialis pills. | No |

One challenge associated with labeling the dataset was that some of the tweets did not contain enough information to be classified easily as selling or not selling drugs. For example, "200 حبوب سايتوتيك"[1] translated as "Cytotec200mg." Some tweets did not include words such as "sell," "provide," "have," "discount code," "price," or "supply" to suggest the purpose is to sell drugs. In these cases, we checked the user's account and timeline: If most of the user's tweets related to selling drugs, then we labeled any ambiguous tweets from that account as selling drugs and vice-versa.

*3.3.2 Monitoring legal and illegal drug sales.* Based on the classification process, we identified 209 positive tweets (tweets selling drugs) posted by 144 users. As it would have been difficult to establish the legality/illegality of the drugs across all Arab countries, we worked with a pharmacist to manually label each tweet according to the legality of selling any drug referenced in a tweet in the United States. We also labeled legal drugs requiring a prescription in the United States but being sold without one on Twitter as illegal (Table 3). We then divided this residual dataset into a 70% training set and a 30% testing set.

Table 3. Examples of tweets with legal/illegal class labels

| Tweet | Label |
|---|---|
| Ubervita w700 is the most popular weight loss pill in states. The offer is valid until Oct. 19, 2016. | Legal |
| Birth control pills (12 SAR). I think it is not expensive. | Illegal |

One challenge with identifying the legality of a drug is its strength. We found that many sellers did not provide this information. Moreover, 15% of the tweets posted with the purpose of selling drugs did not mention a drug by name. Instead of referencing the name of the drug, some sellers provided an image of it and/or a description of its intended use. For instance, "xxxxxxxxxحبوب لزيادة الطول للرجل والمرأة ٧ سم خلال ٩٠ يوم، اطلبها الان على الرقم "[2] translated as "Pills for increasing a woman's and man's height 7 cm in 90 days. Order it now on number xxxxxxxxx." In addition, in some tweets, the drugs referenced were not identified by name and/or were not known to the pharmacist we had engaged. In other tweets, no drug was referred to by name; instead, a pharmaceutical company was mentioned.

**4. Results and discussion**

We found that 66% of the tweets in our dataset were selling legal drugs and 34% were selling illegal drugs and/or selling legal drugs illegally. We evaluated the performance of the SVM classifier based on a receiver operating characteristic (ROC)—a technique for visualizing, organizing, and selecting classifiers based on their performance. Compared with an accuracy or error rate, an ROC provides a richer measure of classification performance [22]. An area of 1 represents a perfect test; an area of .5 represents a worthless test [23]. We used the ROC to determine how well our SVM classifier could separate positive examples from negative examples.

---

[1] https://twitter.com/cytoteccytotec/status/827802647130423296
[2] https://twitter.com/albarari001/status/788111244121300992/

Further, we used the ROC to identify the best threshold for separating these two kinds of examples from each other using this criterion. The area under the curve (AUC) is .98 (Figure 1). We applied 10-fold cross-validation in our evaluation of the classifiers. We compared the accuracy of four classification algorithms and found that SVM achieved 96%, DT 94%, NB 94%, and RF 95%. We applied the same process to evaluating the performance of the classifiers in detecting whether a tweet is selling illegal drugs, selling legal drugs legally, or selling illegal drugs illegally. We achieved 82% accuracy with SVM, 75% with DT, 85% with NB, and 76% with RF (Figure 2).

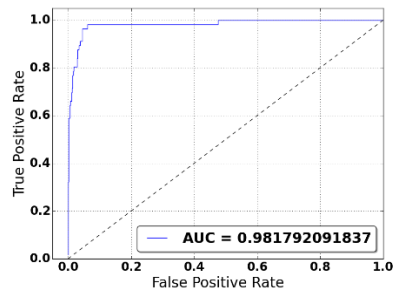
Fig. 1. Receiver operating characteristic (ROC).

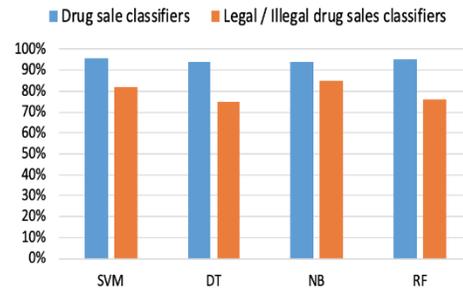
Fig. 2. Accuracy of classifiers.

## 5. Conclusions and future work

In this paper, we proposed a text-classification approach to monitoring drug selling via Arabic-language tweets. We determined whether tweets were or were not selling drugs. Then, for those tweets selling drugs, we determined whether the drugs offered were legal or illegal drugs. Based on our results, SVM and NB performed better than the other classifiers. We intend to work on building a framework drawing on more features (e.g., drug strength) and more tweets to predict the legality of the drugs offered for sale in this context. Further, we plan to develop tools that will render the pre-processing of the data as automatic as possible, utilize recommendation systems [24] and use new social media metrics [25][26][27][28].